\documentstyle[12pt,psfig,draft,aas_macros]{nature} 

\newcommand{\cm}{{\rm \, cm}}
\newcommand{\km}{{\rm \, km}}
\newcommand{\yr}{{\rm \, y}}
\newcommand{\au}{{\rm \, AU}}
\def\refnew#1{(\ref{#1})}

\begin{document}
\title{\large \bf Formation of Kuiper Belt Binaries}

\author{Peter Goldreich, Yoram Lithwick and Re'em Sari}

\summary{ It appears that at least several percent of large Kuiper
belt objects are members of wide binaries. Physical collisions are too
infrequent to account for their formation. Collisionless gravitational
interactions are more promising. These provide two channels for binary
formation. In each, the initial step is the formation of a transient
binary when two large bodies penetrate each other's Hill
spheres. Stabilization of a transient binary requires that it lose
energy.  Either dynamical friction due to small bodies or the
scattering of a third large body can be responsible.  Our estimates
favor the former, albeit by a small margin.  We predict that most
objects of size comparable to those currently observed in the Kuiper
belt are members of multiple systems. More specifically, we derive the
probability that a large body is a member of a binary with semi-major
axis of order $a$. The probability depends upon $\sigma$, the total
surface density, $\Sigma$, the surface density of large bodies having
radius $R$, and $\theta_\odot \approx 10^{-4}$, the angle subtended by
the solar radius as seen from the Kuiper belt. For
$(\sigma/\Sigma)R<a< R/\theta_\odot$, the probability is just
$(\Sigma/\rho R)\theta_\odot^{-2}$, the optical depth of the large
bodies divided by the solid angle subtended by the Sun. For
$R<a<r_u\equiv (\sigma/\Sigma)R$, it varies inversely with semi-major
axis and reaches $\sim (\sigma/\rho R)\theta_\odot^{-2}$ at $a \approx
R$. Based on current surveys of the Kuiper belt, we estimate
$\Sigma/\rho \sim 3 \times 10^{-4}\cm$ and $R \sim 100\km$. We obtain
$\sigma/\rho \sim 0.3\cm $ by extrapolating the surface density
deduced for the minimum mass solar nebula.  Rough predictions are:
outside of the critical separation $r_u/a_\odot\sim 3''$, the binary
probability is $\sim 0.3\%$; at separations of $0.2''$, comparable to
current resolving capabilities, it reaches $\sim 5\%$, in
agreement with results from the HST binary survey by Brown.  }

\maketitle

\section{Introduction}

The Kuiper belt\cite{LuJ02} is the best solar system laboratory for
studies of the early stages of runaway accretion. Runaway accretion in
the Kuiper belt terminated when the velocity dispersion of its members
was increased by an as yet undetermined process\cite{MJP02}. Unlike
the asteroid belt, its largest members have suffered little
collisional evolution since.

The discovery that a substantial fraction of its largest members are
in binaries with wide separations and order unity mass
ratios\cite{EKO01,KPG+01,BrT02,TrB02,NSG+02,NSG+02b,VPG02} is the
latest of many surprises provided by the Kuiper belt. Collisions
coupled with tidal evolution, mechanisms that may explain other solar
system binaries, fail to account for the formation of the Kuiper belt
binaries.

The low frequency of collisions among the large Kuiper belt objects\cite{Ste02}
implies that binaries formed by collisionless interactions mediated by
gravity. These were most effective earlier when dynamical friction due
to small bodies limited the velocity dispersion of the large ones.
This situation pertains during runaway accretion.

In the following section \S \ref{sec:prelim}, we outline a simple
model for runaway accretion to set the stage for binary formation. We
are guided by more detailed formalisms implemented in numerical
simulations.  Values of relevant parameters are estimated based on the
numbers and sizes of objects deduced from Kuiper belt surveys, and by
the extrapolation of the surface density in the minimum mass solar
nebula. We estimate the binary formation rate and derive the
semi-major axis distribution in section \S \ref{sec:binary}. In the
final section \S \ref{sec:discuss}, we discuss the observational implications
of our results and mention several open issues.

\section{Preliminaries}
\label{sec:prelim}

In this section, we introduce a simple set of equations to describe
the major processes that occur during the growth of
planetesimals. These equations are abstracted from more complete
treatments in the literature \cite{WeS89,WeS93}. The main
simplification in our approach is the identification of two groups of
bodies, small ones, containing most of the total mass, and large ones,
contributing a small fraction of it.  The latter group, which
dominates the stirring of all bodies, is identified with the currently
observed population of Kuiper belt objects.

Gravitational interactions determine the accretion rates and velocity
dispersions of both large and small bodies. The governing equations
for these processes read:
\begin{equation}
\label{Rdot}
{1 \over \Omega_\odot} {1 \over R}{dR \over dt}
\sim {\Sigma \over \rho R} F_{M-M}+
{\sigma \over \rho R} F_{M-m},
\end{equation}

\begin{equation}
\label{rdot}
{1 \over \Omega_\odot} {1 \over s}{ds \over dt}\sim
{\sigma \over \rho s},
\end{equation}

\begin{equation}
\label{vdot}
{1 \over \Omega_\odot} {1 \over v}{dv \over dt}\sim
{\Sigma \over \rho R} F_{M-M}^2-
{\sigma \over \rho R} F_{M-m}^2,
\end{equation}

\begin{equation}
\label{udot}
{1 \over \Omega_\odot} {1 \over u}{du \over dt}\sim
{\Sigma \over \rho R} F_{M-m}^2-
{\sigma \over \rho s}, 
\end{equation}
where $R$ and $s$ are the radii of the large and small bodies,
$\Sigma$ and $\sigma$ are their surface mass densities, $v$ and $u$
are their velocity dispersions,\footnote{We are making an additional
simplification by using a single velocity dispersion in place of three
that are needed to characterize a triaxial velocity ellipsoid.} $\rho$
is the material density\footnote{We use the same density for the
Kuiper belt objects as for the Sun.} and $\Omega_\odot$ is the
orbital frequency around the Sun.\footnote{In writing equations
\refnew{Rdot}-\refnew{udot} we implicitly assume that $u>v$.}

The expressions $\Sigma\Omega_\odot/\rho R$ and
$\sigma\Omega_\odot/\rho s$ give the kinematic (ignoring gravitational
focusing) collision rates of large bodies onto large bodies and small
bodies onto small bodies; $\sigma\Omega_\odot/\rho R$ is the kinematic
collision rate of small bodies onto a large one multiplied by the mass
ratio $(s/R)^3$. The factors $F_{M-M}$ and $F_{M-m}$ in equation
\refnew{Rdot} account for the enhancements of the physical collision
rates due to gravitational focusing by large bodies of the
trajectories of incoming large and small bodies respectively. Cross
sections for large deflections exceed those for physical collisions
when gravitational focusing is significant. The factors $F_{M-M}^2$
and $F_{M-m}^2$ that appear in equations \refnew{vdot} and
\refnew{udot} are the ratios of those cross sections to the
geometrical ones. Appropriate expressions for $F_{M-m}$ read:
\begin{equation}
F_{M-m}\sim \cases {  1                  &    $v_{esc}<u$     \cr
                  (v_{esc}/u)^2      &    $v_H<u<v_{esc}$  \cr
                  (v_{esc}^2/v_H u)  &    $\theta_\odot^{1/2}v_H<u<v_H$             }
\end{equation}
where $\theta_\odot \approx 10^{-4}$ is the angle subtended by the
solar radius as seen from the Kuiper belt, $v_{esc}$ is the escape
velocity from the large bodies and $v_H$ is their Hill velocity
defined in terms of the Hill radius,
\begin{equation}
R_H\sim \left({M\over M_\odot}\right)^{1/3}a_\odot\sim {R \over \theta_\odot},
\end{equation}
as
\begin{equation}
v_H\sim \Omega_\odot R_H.
\end{equation}
The escape velocity is related to the Hill velocity by 
\begin{equation}
v_{esc}\sim (G\rho)^{1/2}R\sim v_H/\theta_\odot^{1/2}.
\end{equation}
$F_{M-M}$ is obtained by replacing $u$ with $v$ in the above expressions
and in their ranges of validity. 

Each of equations \refnew{Rdot}-\refnew{udot} has a simple
interpretation. In equation \refnew{Rdot}, the two terms on the rhs
account for the radius growth of the large bodies by the accretion of
large and small bodies.  The single term on the rhs of equation
\refnew{rdot} describes how the radii of small bodies increase as the
result of coalescence under conditions of negligible gravitational
focusing. In equation \refnew{vdot}, the first term describes the
viscous stirring of large bodies due to mutual gravitational
deflections, whereas the second term accounts for the damping of the
large bodies' velocity dispersion by dynamical friction resulting from
their gravitational interactions with the small bodies. Equation
\refnew{udot} shows how the velocity dispersion of the small bodies
evolves under viscous stirring by large bodies and damping due to
collisions between small bodies.

Values of the dimensionless parameters $\Sigma/\sigma\ll 1$ and
$\theta_\odot$ determine the appropriate regime of runaway
accretion. At the Kuiper belt $a_\odot\approx 40\au$ so
$\theta_\odot\approx 10^{-4}$. Based on Kuiper belt
surveys\cite{TJL01,GKN98,ChB99} which detect only the large bodies, we
estimate $\Sigma/\rho\sim 3\times 10^{-4}$cm and
$R\approx 100\km$. Extrapolating the minimum mass solar nebula surface
density to $40$~AU, yields $\sigma/\rho \sim 0.3\cm$. The resulting
ratio of $\Sigma/\sigma\sim 10^{-3}$ is compatible with Kuiper belt
simulations\cite{KeL98,KeL99,Ken02}. In the rest of this paper we
assume that
\begin{equation}
\theta_\odot< {\Sigma \over \sigma} < 1.
\end{equation}

In this regime of runaway accretion, the rates of gravitational
stirring and dynamical friction acting on the large bodies are much
greater than the large bodies' growth rates. Moreover, provided the
radii of the small bodies satisfy
\begin{equation}
\label{rsol}
s>{\Sigma \over \sigma} R \sim 0.1\km,
\end{equation}
their growth rates are negligible compared to those of the large
bodies and collisional damping of their velocities is unimportant on
the timescale of viscous stirring.\footnote{We assume that the
inequality in equation \refnew{rsol} always applies.} Under these
approximations, dynamical friction on the large bodies balances their
viscous stirring, and the stirring of the small bodies proceeds with
negligible collisional losses. Equations \refnew{Rdot}-\refnew{udot}
yield
\begin{equation}
{v\over v_H}\sim {1\over \theta_\odot}
\left(\Sigma\over\sigma\right)^{3/2} \sim 0.3
\end{equation}
and
\begin{equation}
{u\over v_H}\sim
\left({1\over\theta_\odot}{\Sigma\over\sigma}\right)^{1/2} \sim 3.
\end{equation}
Thus, the velocity ratio
\begin{equation}
{u\over v}\sim \theta_\odot^{-1/2}
{\sigma\over\Sigma} \sim 10.
\end{equation}
These results are in accord with the most recent simulations\cite{Ken02}.

Given this solution, the balanced rates of viscous stirring and
dynamical friction for large bodies satisfy
\begin{equation}
\left. -{1\over v}{dv\over dt} \right|_{df} =
\left.  {1\over v}{dv\over dt} \right|_{vs} 
\sim {\sigma\over \rho R} \left(\sigma\over \Sigma\right)^2\Omega_\odot.
\label{eq:vdotss}
\end{equation}
Large bodies grow predominantly by accreting small bodies; The small
bodies' velocity dispersion evolves mainly by viscous stirring
provided by the large bodies. These two processes proceed at equal
rates:
\begin{equation}
{1\over R}{dR\over dt}\sim {1\over u}{du\over dt}\sim 
{\sigma\over \rho R}\left(\sigma\over \Sigma\right)\Omega_\odot.
\end{equation}

\section{Binaries}
\label{sec:binary}

Because $v<\Omega_\odot R_H$, the gravity of a large body
significantly deflects other large bodies at separations smaller than
$R_H$, whereas small bodies are only affected at
separations\footnote{We assume that the small bodies are on hyperbolic
orbits with respect to the large body.} smaller than
\begin{equation}
r_u\sim {GM\over u^2}\sim {\sigma\over \Sigma}R.
\end{equation}
Since $v_H<u<v_{esc}$, we have $R<r_u<R_H$. 

For separations $r<r_u$, the velocity dispersion of the small bodies
is
\begin{equation}
\left(GM\over r\right)^{1/2}\sim \left(r_u\over r\right)^{1/2}u.
\end{equation}
In addition, since the impact parameter for a small body to arrive at
radius $r$ is $ b\sim \left(r_u r\right)^{1/2}$, the continuity
equation implies that the small bodies' number density is enhanced by
a factor
\begin{equation}
{b^2 u \over s^2 u(r_u/s)^{1/2}} \sim \left( r_u \over s\right)^{1/2}.
\end{equation}

\subsection{binary formation}

We identify two distinct channels for binary formation. Both begin
when two large bodies penetrate each other's Hill spheres. This occurs
at a rate
\begin{equation}
{\Sigma\over M} R_H^2\Omega_\odot\sim {\Sigma\over \rho R}
\left(1 \over \theta_\odot\right)^2 \Omega_\odot \sim 10^{-4} \yr^{-1},
\label{eq:transrate}
\end{equation}
per large body. Stabilization of such a transient binary requires
energy loss on timescale $\Omega_\odot^{-1}$. This can be achieved either by
dynamical friction from small bodies or by interaction with a
third large body.

Dynamical friction from small bodies during time
$\Omega_\odot^{-1}$ results in a fractional energy loss
\begin{equation}
{\Delta E\over M v_H^2}\sim {\sigma\over \rho R}
\left(\sigma\over\Sigma\right)^2
 \sim 0.03 .
\label{eq:fracE}
\end{equation}
This is also the fraction of transient binaries that become
bound. Numerical simulations that verify this assertion are described
in the figure. Thus, this mechanism leads to a binary formation rate
per large body
\begin{equation}
{\cal R}_1\sim \left(\sigma\over\rho R\right)^2\left(\sigma\over\Sigma\right)
\left(1\over \theta_\odot\right)^2\Omega_\odot 
\sim 3\times 10^{-6} \yr^{-1}  .
\end{equation}

\begin{figure}[b!]
\centerline{\hbox{\psfig{figure=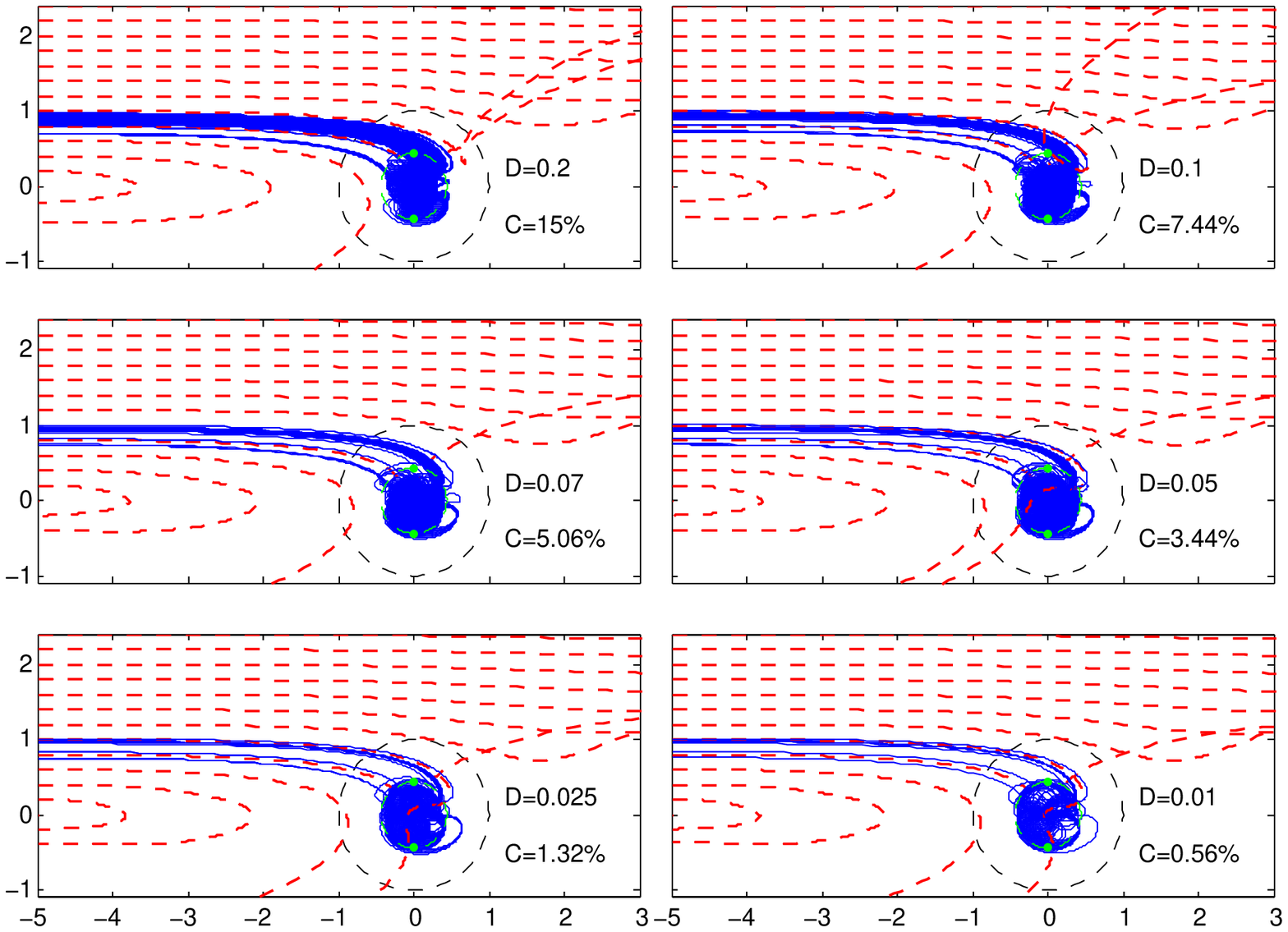}}}
\caption{Numerical simulations of binary formation by dynamical
friction. Two equal mass bodies approach each other on initially
circular orbits.  We integrate the equations of motion including
dynamical friction under the Hill approximation in a frame rotating at
the average mean motion. The center of mass is fixed at (0,0). We plot
only the trajectories of one of the bodies; the trajectories of the
other body are related by reflection through the origin.  Dashed red
trajectories did not form binaries, and solid blue ones did. The axes
are in units of $a_\odot(M/M_\odot)^{1/3}$.  The Lagrange points of
the corresponding restricted 3-body problem are marked by green dots
and the Hill sphere by a dashed green circle. Dynamical friction is
modelled as a force antiparallel and proportional to the velocity
measured relative to the Keplerian shear. Every capture orbit is
plotted but only a sample of the non-capture orbits are shown. It is
evident that capture is only possible within distinct ranges of impact
parameters, whose widths are roughly proportional to the drag.  The
symbol $C$ denotes the ratio of the linear measure of the impact
parameters that result in capture to $a_\odot(M/M_\odot)^{1/3}$. The
symbol $D$ denotes the fractional decrease in velocity due to drag
over a time $\Omega_\odot^{-1}$, essentially half the rhs of equation
\refnew{eq:fracE}. Clearly $C\approx D$ to better than a factor of
two. This verifies our assertion that the fractional energy loss over
time $\Omega_\odot^{-1}$ is also the fraction of transient binaries
that become bound. A sample of similar calculations starting from
orbits with finite velocity $v<v_H$ gives similar results.}
\end{figure}

The probability that a third large body joins the Hill sphere of a
transient binary during its lifetime is
\begin{equation}
{\Sigma\over \rho R}\left(1\over\theta_\odot \right)^2 \sim 3 \times 10^{-3}.
\end{equation}
A significant fraction of these triplets will result in a bound
binary. Therefore, the binary formation rate per large body via this
channel is
\begin{equation}
{\cal R}_2 \sim \left(\Sigma\over\rho R\right)^2
\left(1\over \theta_\odot\right)^4\Omega_\odot
 \sim 3\times 10^{-7} \yr^{-1} .
\end{equation}
With our parameters the ratio
\begin{equation}
{{\cal R}_2\over{\cal R}_1}\sim \left(\Sigma\over\sigma\right)^3
\left(1\over \theta_\odot\right)^2 \sim 0.1
\end{equation}
is smaller than unity. Thus from here on we take ${\cal R}={\cal R}_1$. 

The ratio of ${\cal R}$ to $R^{-1}dR/dt$ is
\begin{equation}
{\sigma\over \rho R} \left(1\over \theta_\odot\right)^2 \sim 3,
\end{equation}
so binaries form at a rate comparable to the large bodies' growth
rate.

\subsection{binary fraction and semi-major axis distribution}

Let $p(a)$ be the differential probability distribution of finding a
large body in a binary with semi-major axis $a$. In steady state\footnote{
By steady state we mean during a time over which $R$ is sensibly constant.}
\begin{equation}
a~p(a)\sim {{\cal R}\over a^{-1}|da/dt|}
\end{equation}
For $r_u<a<R_H$, $a^{-1}da/dt$ equals the energy decay rate due to
dynamical friction given by equation \refnew{eq:vdotss}. Thus
\begin{equation}
a~p(a)\sim {\Sigma\over \rho R}
\left(1\over \theta_\odot\right)^2 \sim 3 \times 10^{-3},
\end{equation}
which is independent of $a$.

For $R<a<r_u$, two competing effects come into play. The number
density of small bodies is enhanced by a factor $(r_u/a)^{1/2}$ and
their velocity dispersion is increased by the same factor.  Therefore
$a^{-1}|da/dt|$ is reduced by a factor $a/r_u$ in comparison to its
constant value for $r_u<a<R_H$. Thus in this interval of semi-major
axis
\begin{equation}
a~p(a)\sim {\sigma\over \rho R}
\left(1\over \theta_\odot\right)^2{R\over a} \sim 3 \left(R\over a\right),
\end{equation}
which increases inwards as $a^{-1}$.

The timescale for a binary to spiral in until contact is achieved is
equal to $R(dR/dt)^{-1}$. Contact occurs on the same time
scale as that during which large bodies grow by accretion of small
ones. Thus, if we define a critical separation
\begin{equation}
a_{crit}\sim {\sigma\over \rho}{1\over \theta_\odot^2} \sim 300\km,
\end{equation}
inside of which $a~p(a)>1$, then accretion by binary inspiral might
make a substantial contribution before $R$ reaches $a_{crit}$. As long
as $R<a_{crit}$, we may also expect to find more than one companion to
a given large body. However, both these assertions are uncertain since
they require the stability of clusters of high multiplicity.  For our
parameters, $R\sim a_{crit}$.

Exchange reactions, in which the lighter member of a binary is
replaced by a heavier body which passes through the system, are rare
occurrences. The rate at which large bodies pass through an existing
binary of semimajor axis $r_H$ is given by equation
\refnew{eq:transrate}. This is smaller than the rate of orbital decay.
For smaller semimajor axes, even fewer large bodies pass through
during an orbital decay time.

\section{Discussion}
\label{sec:discuss}

We propose that the wide binaries observed in the Kuiper belt formed
during runaway accretion. A fraction of the large bodies that entered
each other's Hill spheres became bound as the result of energy lost to
small bodies by dynamical friction.\footnote{Interaction with a third
large body was a less important channel for binary formation.} The
time scale for a large body to become bound to a similar companion was
\begin{equation}
\left(\rho R \over \sigma\right)^2
 \left(\Sigma\over\sigma\right)
{\theta_\odot^2\over \Omega_\odot} \sim 3 \times 10^5 \yr, 
\end{equation} 

Further dynamical friction hardened the binaries. The timescale to
achieve contact was that during which isolated large bodies grew;
\begin{equation}
{\rho R\over \sigma}\left(\Sigma\over\sigma\right){1\over \Omega_\odot}
\sim 10^6\yr.  
\end{equation}
Initially the inspiral was at a constant rate, but it slowed down
inside $r_u$ where the small bodies' number density and velocity
dispersion were enhanced above their background values. We deduce that
the probability that a large body is part of a binary with angular
separation greater than $r_u/a_\odot\sim 3''$ is
\begin{equation}
{\Sigma\over \rho R} \left(1\over \theta_\odot\right)^2 \sim 3\times 10^{-3}.
\end{equation}
Inward of $r_u$, the binary probability per logarithmic interval of
semimajor axis increases inversely with semimajor axis. Close to
contact, the probability exceeds unity for $R<a_{crit}\sim 100\km$,
which implies that systems with higher multiplicities would exist if
such are stable. For the resolution of the HST survey $\sim 0.2''$, we
predict a binary fraction of about $5\%$, roughly compatible with
observations (Brown, private communication). Our prediction that close
binaries are common could be tested by monitoring the brightness of
Kuiper belt objects for evidence of eclipses and/or fast rotation. The
latter is a consequence, but not a unique signature, of binary
mergers.

In our analysis we estimate the surface density ratio of large to
small bodies from the observational census of Kuiper belt objects
together with an extrapolation of the surface density of the minimum
mass solar nebula. It would be an improvement to have a theoretical
understanding of how this ratio evolves during runaway
accretion. Also, we elaborated the binary evolution scenario with a
given set of parameters, specifically $\theta_\odot$ and $\Sigma
/\sigma$. Different scalings that apply for other parameter regimes
remain to be worked out.

Additional predictions suitable for testing by future observations of
Kuiper belt binaries could be made by extending our model in a variety
of ways. Relaxing the restriction to two mass groups would allow us to
investigate the mass ratio distribution of binary components and its
dependence on semi-major axis. We predict the formation of multiple
systems for bodies with $R<a_{crit}$. Numerical experiments can
establish what types of systems survive disruption due to mutual
interactions of their members. We could also predict eccentricity and
inclination distributions. This would require a more detailed look at
the anisotropy of the velocity distribution of small bodies,
especially for $a<r_u$. Anisotropy leads to tensorial dynamical
friction in which the force is not antiparallel to the
velocity. Binney\cite{Bin77} has investigated a related problem, the
evolution of the orbits of galaxies in anisotropic clusters. We can
already say that the eccentricities are likely to be large.  They
begin large following binary capture. Moreover, since the magnitude of
the dynamical friction force increases outward for $a<r_u$, they
should remain large in this region.

Simple numerical simulations of the Hill problem that incorporate
dynamical friction would help pin down the dependence of the capture
probability on the velocity dispersion and on the strength of
dynamical friction. Computations involving three massive bodies are
needed for evaluation of capture probabilities via that channel. Three
body capture cross sections have been computed for stars in globular
clusters \cite{HuB83}, but in our regime the Hill sphere must be taken
into account.

The two largest known Kuiper belt objects, Pluto and Charon, form a
binary. Collision coupled with tidal evolution is the generally
accepted mode of formation for this system. Our scenario for binary
formation provides another candidate. It may also apply to the
formation of some binaries in the asteroid belt. 

In closing, we caution that our numerical results are crude. Dynamical
processes have been estimated to order of magnitude. We have
ruthlessly discarded numerical coefficients and logarithmic factors.
More credence should be accorded to the functional dependences of our
results than to the precise numerical predictions.

\bibliographystyle{natureedo}
\bibliography{kuiper}

\end{document}